\PassOptionsToPackage{unicode}{hyperref}
\PassOptionsToPackage{hyphens}{url}
\PassOptionsToPackage{dvipsnames,svgnames,x11names}{xcolor}
\documentclass[
  12pt,
]{article}

\usepackage{amsmath,amssymb}
\usepackage{setspace}
\usepackage{iftex}
\ifPDFTeX
  \usepackage[T1]{fontenc}
  \usepackage[utf8]{inputenc}
  \usepackage{textcomp} 
\else 
  \usepackage{unicode-math}
  \defaultfontfeatures{Scale=MatchLowercase}
  \defaultfontfeatures[\rmfamily]{Ligatures=TeX,Scale=1}
\fi
\usepackage{lmodern}
\ifPDFTeX\else  
\fi
\IfFileExists{upquote.sty}{\usepackage{upquote}}{}
\IfFileExists{microtype.sty}{
  \usepackage[]{microtype}
  \UseMicrotypeSet[protrusion]{basicmath} 
}{}
\makeatletter
\@ifundefined{KOMAClassName}{
  \IfFileExists{parskip.sty}{
    \usepackage{parskip}
  }{
    \setlength{\parindent}{0pt}
    \setlength{\parskip}{6pt plus 2pt minus 1pt}}
}{
  \KOMAoptions{parskip=half}}
\makeatother
\usepackage{xcolor}
\usepackage[top=1in,bottom=1in,left=1in,right=1in]{geometry}
\ifx\paragraph\undefined\else
  \let\oldparagraph\paragraph
  \renewcommand{\paragraph}[1]{\oldparagraph{#1}\mbox{}}
\fi
\ifx\subparagraph\undefined\else
  \let\oldsubparagraph\subparagraph
  \renewcommand{\subparagraph}[1]{\oldsubparagraph{#1}\mbox{}}
\fi

\usepackage{longtable,booktabs,array}
\usepackage{calc} 
\usepackage{etoolbox}
\makeatletter
\patchcmd\longtable{\par}{\if@noskipsec\mbox{}\fi\par}{}{}
\makeatother
\IfFileExists{footnotehyper.sty}{\usepackage{footnotehyper}}{\usepackage{footnote}}
\makesavenoteenv{longtable}
\usepackage{graphicx}
\makeatletter
\def\maxwidth{\ifdim\Gin@nat@width>\linewidth\linewidth\else\Gin@nat@width\fi}
\def\maxheight{\ifdim\Gin@nat@height>\textheight\textheight\else\Gin@nat@height\fi}
\makeatother
\setkeys{Gin}{width=\maxwidth,height=\maxheight,keepaspectratio}
\makeatletter
\def\fps@figure{htbp}
\makeatother

\usepackage{booktabs}
\usepackage{longtable}
\usepackage{array}
\usepackage{multirow}
\usepackage{wrapfig}
\usepackage{float}
\usepackage{colortbl}
\usepackage{pdflscape}
\usepackage{tabu}
\usepackage{threeparttable}
\usepackage{threeparttablex}
\usepackage[normalem]{ulem}
\usepackage{makecell}
\usepackage{xcolor}
\usepackage{booktabs}
\usepackage{float}
\usepackage{dcolumn}
\usepackage{eurosym}
\makeatletter
\@ifpackageloaded{caption}{}{\usepackage{caption}}
\AtBeginDocument{
\ifdefined\contentsname
  \renewcommand*\contentsname{Table of contents}
\else
  \newcommand\contentsname{Table of contents}
\fi
\ifdefined\listfigurename
  \renewcommand*\listfigurename{List of Figures}
\else
  \newcommand\listfigurename{List of Figures}
\fi
\ifdefined\listtablename
  \renewcommand*\listtablename{List of Tables}
\else
  \newcommand\listtablename{List of Tables}
\fi
\ifdefined\figurename
  \renewcommand*\figurename{Figure}
\else
  \newcommand\figurename{Figure}
\fi
\ifdefined\tablename
  \renewcommand*\tablename{Table}
\else
  \newcommand\tablename{Table}
\fi
}
\@ifpackageloaded{float}{}{\usepackage{float}}
\floatstyle{ruled}
\@ifundefined{c@chapter}{\newfloat{codelisting}{h}{lop}}{\newfloat{codelisting}{h}{lop}[chapter]}
\floatname{codelisting}{Listing}

\makeatother
\makeatletter
\@ifpackageloaded{caption}{}{\usepackage{caption}}
\@ifpackageloaded{subcaption}{}{\usepackage{subcaption}}
\makeatother
\ifLuaTeX
  \usepackage{selnolig}  
\fi
\usepackage[]{natbib}
\usepackage{bookmark}

\IfFileExists{xurl.sty}{\usepackage{xurl}}{} 
\hypersetup{
  pdftitle={Leg Drain: Quantifying the Global Redistribution of Football Talent through Multi-National Eligibility},
  pdfauthor={Alexander Lehner; Giovanni Righetto},
  colorlinks=true,
  linkcolor={blue},
  filecolor={Maroon},
  citecolor={Blue},
  urlcolor={Blue},
  pdfcreator={LaTeX via pandoc}}

\title{Leg Drain: Quantifying the Global Redistribution of Football
Talent through Multi-National Eligibility}
\author{Alexander
Lehner\thanks{World Bank Group, Outcomes Department, 1818 H Street, Washington, DC. Email: alehner@worldbank.org. The findings, interpretations, and conclusions expressed in this paper are entirely those of the author and do not necessarily represent the views of the World Bank Group, its Executive Directors, or the countries they represent.} \and Giovanni
Righetto\thanks{University of Bologna, Department of Economics, Piazza Scaravilli 2, 40126 Bologna}}
\date{2026-03-17}

\begin{document}
\maketitle
\begin{abstract}
Brain drain---the emigration of skilled individuals toward higher-wage
economies---is a well-documented phenomenon, yet its aggregate economic
cost remains difficult to quantify because individual productivity is
rarely observed. We offer a novel angle on this measurement challenge by
studying professional football, a global labour market in which every
participant carries a publicly observable, consistently estimated market
value. Using data on over 92,000 professional footballers worldwide from
Transfermarkt, we identify nearly 20,000 players with multi-national
eligibility and compute the implied transfer of human capital between
countries. We find that the resulting ``leg drain'' disproportionately
benefits wealthy European nations---France alone gains over €3 billion
in player value---while African and Caribbean countries bear the largest
losses relative to GDP. Italy is the single largest net loser in
absolute terms, driven by the outflow of players with Italian heritage
to Latin American national teams. A gravity model of bilateral flows
reveals that former colonial ties are among the strongest predictors of
leg drain intensity: countries with a colonial relationship to a major
European footballing nation lose significantly more player value, even
after controlling for population and income. These findings provide a
transparent, quantifiable analogue to the broader brain drain debate and
highlight how historical institutional links continue to shape global
talent redistribution.


\end{abstract}

\setstretch{1.5}
\section{Introduction}\label{introduction}

The emigration of skilled individuals from poorer to richer
countries---commonly termed ``brain drain''---has been a persistent
feature of globalisation over the past half-century. Beginning with the
foundational contributions of \citet{bhagwati1974education} and
\citet{grubel1966international}, economists have debated both the
magnitude and welfare consequences of high-skill emigration. Subsequent
work has established that brain drain is large, persistent, and
geographically concentrated: \citet{docquier2012globalization} document
that over 20 million tertiary-educated individuals from developing
countries reside in the OECD, with emigration rates exceeding 80 percent
in some small Caribbean and Pacific nations. More recently,
\citet{delogu2018globalizing} show through a dynamic general-equilibrium
framework that liberalising high-skill migration could yield substantial
global welfare gains, while \citet{burzynski2022climate} demonstrate
that climate change will intensify these flows, disproportionately
affecting the poorest countries. Yet despite decades of research, a
fundamental measurement challenge remains: the economic value of
individual workers is seldom directly observable, making it nearly
impossible to compute how much human capital a sending country actually
loses when its citizens emigrate
\citep{clemens2011economics, docquier2012globalization}.

This measurement gap has shaped---and limited---the entire brain drain
debate. Theoretical predictions depend critically on the magnitude of
the human capital loss: moderate outflows may generate beneficial
``brain gain'' incentives for education
\citep{beine2011panel, chand2023human}, while large outflows may
impoverish origin countries and widen global inequality
\citep{biavaschi2020taking, bocquier2024within}. Empirical work has
therefore relied on indirect proxies---educational attainment,
occupation categories, bilateral migration stocks
\citep{ozden2011where, grogger2011income, abel2019bilateral}---none of
which attach a market price to the individual worker. The result is that
we know \emph{who} moves but not precisely \emph{how much} their
departure costs the sending country.

We propose to overcome this impasse by studying one labour market where
individual valuations are both public and consistently estimated across
the globe: professional football. The website Transfermarkt, the world's
most widely referenced player valuation platform, assigns crowd-sourced
market values to over 800,000 players worldwide. These valuations have
been shown to correlate strongly with actual transfer fees and to
reflect observable player quality
\citep{herm2014when, muller2017beyond, peeters2018testing}. Crucially,
Transfermarkt data spans every tier of professional football in
virtually every country---enabling a global analysis that is rare even
in the broader migration literature.

Our key insight is simple. When a dual-citizenship footballer chooses to
represent a national team other than his country of origin, he
effectively ``transfers'' his human capital from one country to
another---at least in the symbolic and sporting sense that national
teams embody a country's talent pool. We call this phenomenon \emph{leg
drain}, in analogy to brain drain. Unlike brain drain, however, leg
drain can be precisely quantified: every player carries an observable
market value, and every national team choice is publicly recorded.
Football thus provides a rare setting in which the three ingredients
needed for a complete accounting of skill flows---individual
identification, bilateral assignment, and cardinal valuation---are
simultaneously available.

Using data on 92,643 professional footballers worldwide, we identify
19,956 players with multi-national eligibility who play for a national
team different from at least one of their citizenships. We compute
bilateral flows of player value between countries and merge these with
World Bank indicators on population and GDP to produce normalised
measures of leg drain intensity.

Three findings stand out. First, leg drain is massive in absolute terms:
the total value of players ``redirected'' through dual citizenship
exceeds €20 billion. Second, the flows are highly asymmetric: a handful
of European countries---France, England, the Netherlands, and
Germany---are the primary beneficiaries, while African, Caribbean, and
Central American nations bear the largest losses. Third, colonial
history is a powerful predictor of these flows: former colonies of
France, the United Kingdom, Belgium, and Portugal lose significantly
more player value to their former colonisers than other countries, even
after controlling for income and population. This pattern mirrors a
well-established finding in the broader migration literature, where
colonial and linguistic ties are among the strongest predictors of
bilateral migration corridors
\citep{beine2011diasporas, head2014gravity, beine2015parsons, bertoli2015visa}.

Our contribution is threefold. First, we offer a \emph{methodological}
contribution by demonstrating that the football labour market---where
individual identification, country assignment, and cardinal valuation
are simultaneously observable---can serve as a transparent, globally
comparable laboratory for studying the forces that shape talent
redistribution. While national-team representation does not entail a
change in a player's club employment or residence, the bilateral
patterns it generates are shaped by the same colonial, linguistic, and
institutional links that drive labour migration. Unlike conventional
approaches that infer human capital from education or occupation, our
setting provides direct market valuations for each individual, enabling
precise bilateral accounting. Second, we make a \emph{descriptive}
contribution by constructing the first comprehensive dataset of leg
drain flows---covering 201 countries, over 2,300 bilateral corridors,
and more than €20 billion in player value---and documenting the striking
asymmetry between net gainers and losers. Third, we make an
\emph{analytical} contribution by embedding these flows in a gravity
framework and showing that colonial ties remain a dominant determinant
of talent redistribution, with effect sizes that are, if anything,
larger than those found in the general migration literature
\citep{beine2011diasporas, docquier2016emigration}. Together, these
results highlight how historical institutional links continue to shape
the global distribution of human capital---not only in the abstract, but
in a domain where the consequences are publicly visible every time a
national team takes the field.

The remainder of this paper is organised as follows. Section \ref{data}
describes the data. Section \ref{descriptive-evidence} presents
descriptive evidence on leg drain flows. Section
\ref{normalised-measures-and-regression-analysis} introduces normalised
measures and a gravity model of bilateral flows. Section
\ref{sec-colonial} discusses the colonial dimension in detail. Section
\ref{discussion-and-conclusion} concludes.

\section{Data}\label{data}

\subsection{Transfermarkt}\label{transfermarkt}

Our player-level data come from the \emph{football-datasets}
repository,\footnote{Accessible via
  \url{https://github.com/salimt/football-datasets}.} which compiles
Transfermarkt profiles for 92,643 professional footballers worldwide
across all leagues and tiers, last updated in October 2025. For each
player, we observe the latest market value (in euros), primary and
secondary citizenship, age, position, and club affiliation.

Market values on Transfermarkt are determined through a structured
community process: registered users propose valuations based on
performance, age, contract duration, and other factors, and moderators
aggregate these into a consensus estimate. Research has validated these
values as reliable proxies for true market prices
\citep{herm2014when, muller2017beyond, peeters2018testing}.

\subsection{Identifying Leg Drain}\label{identifying-leg-drain}

We define a player as having \emph{multi-national elgibility} if
Transfermarkt records two distinct citizenships for him. The citizenship
field in the source data lists multiple nationalities separated by a
double-space delimiter (e.g., ``Belgium DR Congo''). We parse the first
entry as the \emph{primary} citizenship---corresponding to the national
team the player represents---and the second as the \emph{secondary}
citizenship, indicating an alternative national affiliation. We
interpret the secondary citizenship as the ``origin'' country and the
primary citizenship as the ``destination'' country. We verified this
assignment for the 50 highest-valued dual-citizenship players against
their international career records on Transfermarkt. The sample is
dominated by second-generation immigrants who represent the country in
which they were raised, though the direction of the flow varies: some
players born in Europe choose to represent their parents' country of
origin rather than their country of birth. This is, for example,
observable for countries like Algeria, Morocco, or Türkiye. A fourth,
distinct, category of multi-national eligiblity arises from deliberate
passport acquisition, often facilitated by clubs to circumvent non-EU
player quotas. In our data this almost exclusively affects the
interpretation of the numbers for Spain where players who are citizens
of Latin American countries can obtain Spanish citizenship relatively
quickly\footnote{Under \textbf{Article 22 of the Spanish Civil Code},
  nationals of Ibero-American countries (all of Latin America, plus
  Andorra, Philippines, Equatorial Guinea, and Portugal) can apply for
  Spanish citizenship after just \textbf{2 years} of legal residency,
  compared to 10 years for most other nationalities.}. Even though these
specific dual citizenships do not reflect geniune diaspora ties, they do
reflect the modern reality of global labour markets and even amplify the
phenomenon of leg drain by giving players additional options to switch
between nationalities. Furthermore, these types of dual citizenships
illustrate how institutional legacies---in this case, preferential
naturalisation rules rooted in colonial and linguistic links---continue
to shape the redistribution of talent.

Of the 92,643 players in our dataset, 19,956 (21.5\%) hold dual
citizenship, spanning 201 countries. These dual citizens account for
€20.70 billion in market value---39.2\% of the total €52.87
billion---despite comprising only about a fifth of all players.

\subsection{World Bank Indicators}\label{world-bank-indicators}

We match each country to World Bank World Development Indicators (WDI)
for 2024 (with 2023 as fallback for missing values)
\citep{WorldBankWDI}: total population, GDP (current USD), and GDP per
capita. This allows us to normalise leg drain flows by country size and
wealth. Country names from Transfermarkt are matched to ISO 3166-1 codes
using the \texttt{countrycode} package, with manual overrides for
entities such as England, Scotland, and Wales (mapped to GBR for WDI
purposes).

\section{Descriptive Evidence}\label{descriptive-evidence}

\subsection{Aggregate Flows}\label{aggregate-flows}

\begin{table}[!h]
\centering
\caption{\label{tab:aggregate-flows}Aggregate leg drain flows (€ millions). Panel A: top 10 by gross value lost and gained. Panel B: top 10 by net flows.}
\centering
\resizebox{\ifdim\width>\linewidth\linewidth\else\width\fi}{!}{
\begin{tabular}[t]{lrlr>{\centering\arraybackslash}p{1cm}lrlr}
\toprule
\multicolumn{4}{c}{\textbf{Panel A: Gross flows}} & \multicolumn{1}{c}{\textbf{ }} & \multicolumn{4}{c}{\textbf{Panel B: Net flows}} \\
\cmidrule(l{3pt}r{3pt}){1-4} \cmidrule(l{3pt}r{3pt}){6-9}
\multicolumn{2}{c}{Value lost} & \multicolumn{2}{c}{Value gained} & \multicolumn{1}{c}{ } & \multicolumn{2}{c}{Largest losers} & \multicolumn{2}{c}{Largest gainers} \\
\cmidrule(l{3pt}r{3pt}){1-2} \cmidrule(l{3pt}r{3pt}){3-4} \cmidrule(l{3pt}r{3pt}){6-7} \cmidrule(l{3pt}r{3pt}){8-9}
Country & €m & Country & €m &  & Country & €m & Country & €m\\
\midrule
Italy & 1,767.0 & France & 3,390.5 &  & Italy & -1,369.5 & France & 1,812.9\\
France & 1,577.6 & England & 2,557.6 &  & Spain & -829.7 & England & 1,119.9\\
Spain & 1,521.9 & Argentina & 1,191.0 &  & Nigeria & -818.7 & Argentina & 1,063.1\\
England & 1,437.7 & Brazil & 1,185.8 &  & DR Congo & -628.3 & Brazil & 1,012.2\\
Nigeria & 1,085.5 & Netherlands & 1,087.3 &  & Cote d'Ivoire & -506.5 & Netherlands & 644.1\\
Cote d'Ivoire & 820.8 & Germany & 1,034.2 &  & Cameroon & -504.5 & Morocco & 460.9\\
DR Congo & 794.9 & Spain & 692.2 &  & Ireland & -421.0 & Germany & 460.0\\
Cameroon & 658.5 & Morocco & 607.5 &  & Jamaica & -417.0 & Uruguay & 438.1\\
Ireland & 609.0 & Portugal & 566.1 &  & Ghana & -411.4 & Sweden & 404.7\\
Germany & 574.1 & Belgium & 535.4 &  & Suriname & -400.9 & United States & 266.5\\
\bottomrule
\end{tabular}}
\end{table}

In absolute terms, Italy loses the most value, followed by France and
Spain. On the gaining side, France dominates, acquiring over €3 billion
in player value through dual-citizenship players---a figure that dwarfs
all other countries. England ranks second, followed by Argentina and
Brazil. As noted earlier, the outflow number for Spain is inflated by
Latin American players who become eligible to obtain Spanish citizenship
after two years of residency and thus help clubs circumvent non-EU
player quotas.

\subsection{Mapping Leg Drain}\label{mapping-leg-drain}

\begin{figure}[H]

{\centering \includegraphics{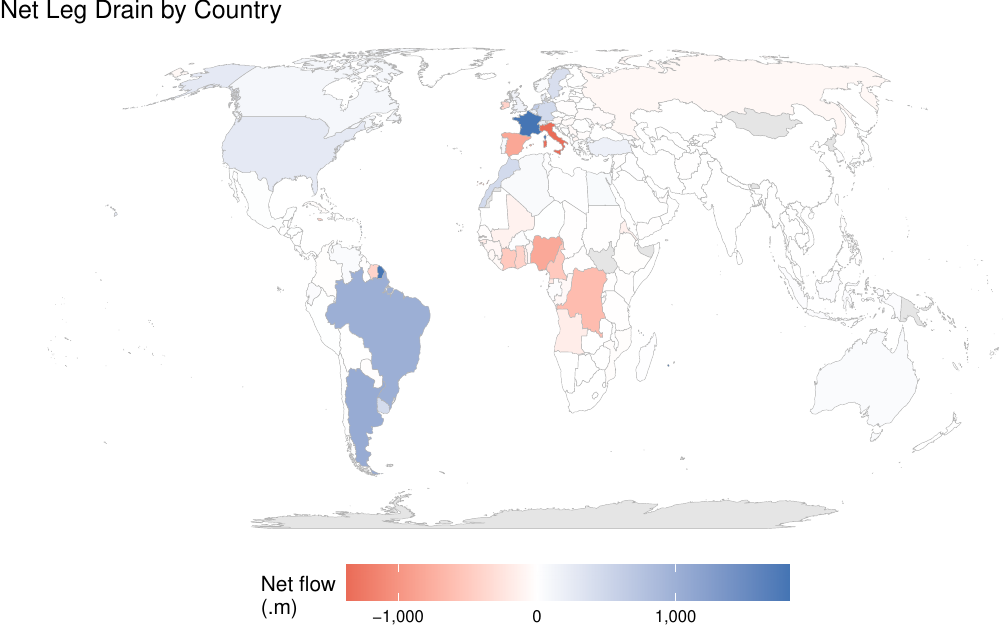}

}

\caption{Net leg drain by country (€ millions). Red indicates net loss;
blue indicates net gain. Countries in grey are not in our sample.}

\end{figure}

Figure 1 maps net leg drain globally. The pattern is stark: Western
European nations appear in blue (net gainers), while Africa and the
Caribbean are in red (net losers). The asymmetry mirrors the broader
geography of global migration and economic inequality.

\subsection{The Best XI
Counterfactual}\label{the-best-xi-counterfactual}

To make the impact of leg drain tangible, we construct a ``Best XI''
comparison for each country. For every national team, we select the
eleven most valuable players and sum their market values. We then repeat
this exercise under a counterfactual in which no dual-citizenship
migration had occurred: dual citizens are reassigned to their origin
country (secondary citizenship), while single-citizenship players remain
unchanged. The difference between these two totals measures how much a
country's best squad gains or loses from leg drain.

\begin{table}[!h]
\centering
\caption{\label{tab:best-xi-tables}Best XI counterfactual analysis. Panel A: countries whose Best XI benefits from leg drain (top 15 by value gained). Panel B: countries whose Best XI is weakened (top 15 by value lost). \% of GDP expresses the Best XI value change as a share of the country's GDP (current USD).}
\centering
\resizebox{\ifdim\width>\linewidth\linewidth\else\width\fi}{!}{
\begin{tabular}[t]{lrrrrrl}
\toprule
Country & Actual (€m) & CF (€m) & Diff. (€m) & Change (\%) & \% of GDP & Top player\\
\midrule
\addlinespace[0.3em]
\multicolumn{7}{l}{\textbf{Panel A: Countries that gain from leg drain}}\\
\hspace{1em}France & 960.0 & 447.0 & 513.0 & 114.8 & 0.0162 & Kylian Mbappé (342229) (€180m)\\
\hspace{1em}Sweden & 459.0 & 177.0 & 282.0 & 159.3 & 0.0467 & Alexander Isak (349066) (€120m)\\
\hspace{1em}Brazil & 835.0 & 560.0 & 275.0 & 49.1 & 0.0126 & Vinicius Junior (371998) (€170m)\\
\hspace{1em}England & 1,050.0 & 795.0 & 255.0 & 32.1 & 0.0069 & Jude Bellingham (581678) (€180m)\\
\hspace{1em}Uruguay & 390.0 & 188.0 & 202.0 & 107.4 & 0.2495 & Federico Valverde (369081) (€130m)\\
\hspace{1em}Argentina & 650.0 & 472.0 & 178.0 & 37.7 & 0.0279 & Alexis Mac Allister (534033) (€100m)\\
\hspace{1em}Morocco & 320.0 & 160.5 & 159.5 & 99.4 & 0.0993 & Achraf Hakimi (398073) (€80m)\\
\hspace{1em}Netherlands & 590.0 & 450.0 & 140.0 & 31.1 & 0.0115 & Ryan Gravenberch (478573) (€75m)\\
\hspace{1em}Germany & 677.0 & 544.0 & 133.0 & 24.4 & 0.0028 & Jamal Musiala (580195) (€140m)\\
\hspace{1em}Belgium & 391.0 & 285.0 & 106.0 & 37.2 & 0.0158 & Loïs Openda (368887) (€50m)\\
\hspace{1em}Wales & 162.0 & 63.2 & 98.8 & 156.3 & 0.0027 & Brennan Johnson (470607) (€40m)\\
\hspace{1em}Türkiye & 282.0 & 187.5 & 94.5 & 50.4 & 0.0070 & Kenan Yıldız (845654) (€50m)\\
\hspace{1em}Switzerland & 240.0 & 162.5 & 77.5 & 47.7 & 0.0083 & Dan Ndoye (365108) (€35m)\\
\hspace{1em}Egypt & 149.3 & 79.5 & 69.8 & 87.8 & 0.0179 & Omar Marmoush (445939) (€75m)\\
\hspace{1em}United States & 269.0 & 206.0 & 63.0 & 30.6 & 0.0002 & Christian Pulisic (315779) (€50m)\\
\addlinespace[0.3em]
\multicolumn{7}{l}{\textbf{Panel B: Countries that lose from leg drain}}\\
\hspace{1em}Cameroon & 201.0 & 608.0 & -407.0 & -66.9 & -0.7636 & Kylian Mbappé (342229) (€180m)\\
\hspace{1em}Ireland & 172.0 & 575.0 & -403.0 & -70.1 & -0.0662 & Jude Bellingham (581678) (€180m)\\
\hspace{1em}Suriname & 24.2 & 351.0 & -326.8 & -93.1 & -7.3991 & Ryan Gravenberch (478573) (€75m)\\
\hspace{1em}Nigeria & 329.0 & 615.0 & -286.0 & -46.5 & -0.1134 & Bukayo Saka (433177) (€150m)\\
\hspace{1em}Guadeloupe & 17.2 & 275.0 & -257.8 & -93.7 & NA & Marcus Thuram (318528) (€75m)\\
\hspace{1em}DR Congo & 127.5 & 355.0 & -227.5 & -64.1 & -0.3206 & Eduardo Camavinga (640428) (€60m)\\
\hspace{1em}Jamaica & 58.0 & 273.0 & -215.0 & -78.8 & -0.9766 & Morgan Gibbs-White (429014) (€50m)\\
\hspace{1em}Martinique & 8.9 & 209.5 & -200.6 & -95.8 & NA & Warren Zaïre-Emery (810092) (€55m)\\
\hspace{1em}Ghana & 195.0 & 395.0 & -200.0 & -50.6 & -0.2430 & Nico Williams (709187) (€70m)\\
\hspace{1em}Equatorial Guinea & 17.4 & 209.2 & -191.9 & -91.7 & -1.5032 & Lamine Yamal (937958) (€200m)\\
\hspace{1em}Italy & 560.0 & 745.0 & -185.0 & -24.8 & -0.0078 & Alexis Mac Allister (534033) (€100m)\\
\hspace{1em}St. Kitts \& Nevis & 1.1 & 177.4 & -176.3 & -99.4 & -15.7053 & Cole Palmer (568177) (€120m)\\
\hspace{1em}Togo & 33.1 & 195.8 & -162.7 & -83.1 & -1.5271 & Cody Gakpo (434675) (€70m)\\
\hspace{1em}Cote d'Ivoire & 305.0 & 463.0 & -158.0 & -34.1 & -0.1814 & Désiré Doué (914562) (€90m)\\
\hspace{1em}Curacao & 18.8 & 169.0 & -150.2 & -88.9 & -4.2177 & Jurriën Timber (420243) (€55m)\\
\bottomrule
\end{tabular}}
\end{table}

\begin{figure}[H]

{\centering \includegraphics{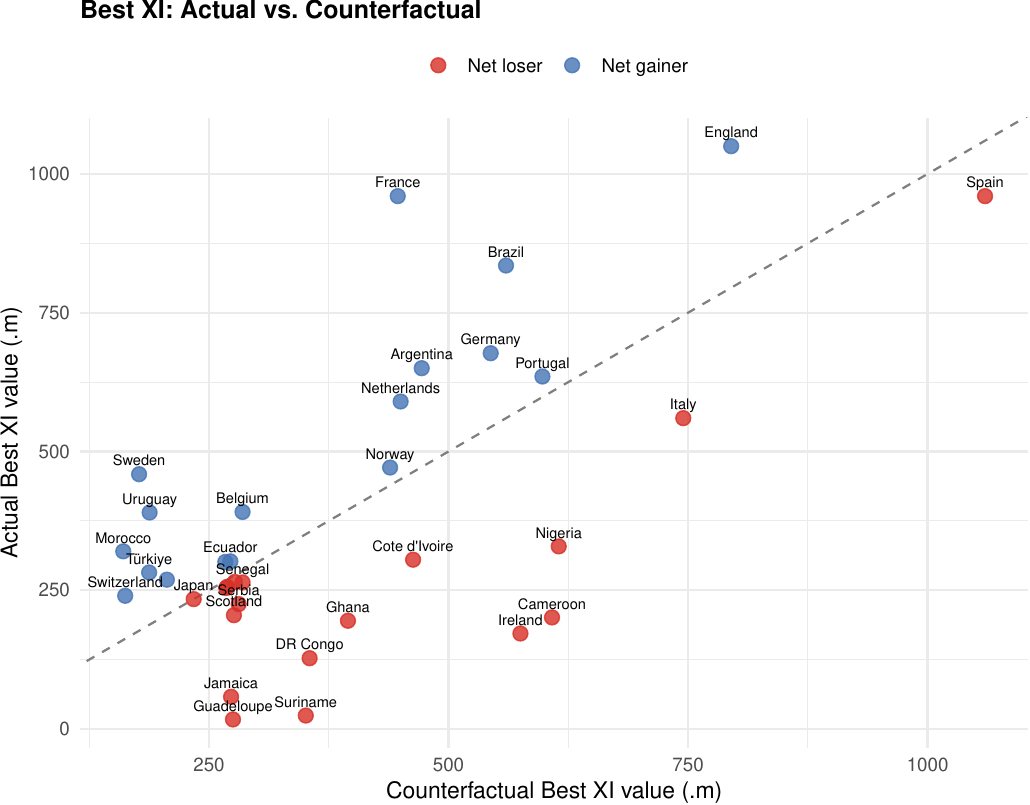}

}

\caption{Best XI squad value under actual rosters vs.~the no-migration
counterfactual. Countries above the 45-degree line benefit from leg
drain; countries below it are weakened.}

\end{figure}

The results are striking. DR Congo is one of the most dramatic cases:
its actual Best XI is worth just €128m, but under the counterfactual it
would command €355m---a squad featuring Romelu Lukaku, Youri Tielemans,
and Christopher Nkunku, among others. Similarly, Jamaica, Guadeloupe,
and Cote d'Ivoire would see their squads transformed. On the winning
side, France, Argentina, Belgium, and Uruguay all field Best XIs that
are substantially more valuable than their counterfactual rosters,
having attracted dual-citizenship talent from across the developing
world. Spain and Italy are notable exceptions among wealthy European
nations: both see their counterfactual squads valued \emph{higher} than
their actual ones. In the case of Italy, this reflects the outflow of
players with ancestral heritage to Latin American national teams. In the
case of Spain, in several cases the dual citizenship does not represent
leg drain, but, as discussed earlier, clubs having an incentive to
facilitate players getting European passports to circumvent non-EU
player quotas, e.g.~Real Madrid with Vinicius Jr.~or Barcelona with
Messi. However, the overall pattern overwhelmingly confirms that leg
drain reshapes the competitive landscape of international football in
systematic and quantitatively significant ways.

\section{Normalised Measures and Regression
Analysis}\label{normalised-measures-and-regression-analysis}

Raw value flows favour large, wealthy countries. To compare leg drain
intensity across countries of different sizes, we construct normalised
measures. When normalised by GDP (Table 2, column ``\% of GDP''), the
picture shifts dramatically. Small Caribbean and African nations---whose
raw losses may appear modest---emerge as the most severely affected.
This parallels findings in the brain drain literature, where small
developing countries suffer disproportionate losses of skilled workers
\citep{docquier2012globalization}.

\subsection{Gravity Model of Bilateral Leg
Drain}\label{gravity-model-of-bilateral-leg-drain}

Country-level regressions aggregate away the bilateral structure of leg
drain. A more appropriate framework is the gravity model, widely used in
the trade and migration literatures, which models flows between
origin--destination pairs as a function of country characteristics and
bilateral linkages.

We construct a bilateral dataset of 1668 origin--destination pairs (all
observed dual-citizenship corridors with available WDI data) and
estimate gravity equations of the form:

\[\text{Flow}_{od} = f(\text{Pop}_o, \text{Pop}_d, \text{GDP/cap}_o, \text{GDP/cap}_d, \text{ColonialTie}_{od})\]

where \(o\) indexes the origin (losing) country and \(d\) the
destination (gaining) country. We estimate this using OLS on log values,
Poisson pseudo-maximum likelihood (PPML), and PPML with destination
fixed effects \citep[see, e.g.,][for a practitioner's
guide]{beinebertoli2016gravity}. The colonial-tie indicator equals one
when the origin country is a former colony of the destination country.

\begin{table}[!h]
\centering
\caption{\label{tab:gravity-regressions}Gravity model of bilateral leg drain flows. Columns (1)--(3): dependent variable is player value in euros. Column (4): dependent variable is player count.}
\centering
\resizebox{\ifdim\width>\linewidth\linewidth\else\width\fi}{!}{
\begin{threeparttable}
\begin{tabular}[t]{lrrrr}
\toprule
 & (1) & (2) & (3) & (4)\\
\midrule
Intercept & 10.360*** (1.486) & 9.889*** (2.448) &  & -4.881*** (1.877)\\
Log pop. (origin) & 0.064 (0.048) & 0.115 (0.075) & 0.160* (0.082) & 0.196*** (0.055)\\
Log pop. (destination) & 0.115** (0.044) & 0.207*** (0.070) &  & 0.127*** (0.044)\\
Log GDP/cap (origin) & -0.023 (0.061) & 0.021 (0.091) & 0.076 (0.095) & 0.108 (0.069)\\
Log GDP/cap (destination) & 0.072 (0.062) & 0.007 (0.082) &  & 0.024 (0.091)\\
Colonial tie (origin to dest.) & 2.146*** (0.496) & 2.214*** (0.243) & 2.096*** (0.224) & 1.770*** (0.447)\\
Observations & 1668 & 2363 & 2341 & 2363\\
Destination FE & No & No & Yes & No\\
Estimator & OLS & PPML & PPML & Poisson\\
\bottomrule
\end{tabular}
\begin{tablenotes}[para]
\item \textit{Note: } 
\item Two-way clustered standard errors (by origin and destination country) in parentheses, following standard practice in the gravity literature. The OLS specification (column 1) uses log(value) as the dependent variable, which drops corridors with zero total player value since log(0) is undefined. PPML and Poisson models (columns 2--4) estimate in levels and therefore retain all corridors, including those where dual-citizen players carry a zero market valuation. * \$p<0.10\$, ** \$p<0.05\$, *** \$p<0.01\$.
\end{tablenotes}
\end{threeparttable}}
\end{table}

The results are consistent across specifications. The colonial-tie
variable is large and highly significant in every column. In the
preferred PPML specification with destination fixed effects (column 3),
a colonial relationship between origin and destination is associated
with a roughly eight-fold increase in the value of player flows
(\(e^{2.096} \approx 8.1\), \(p < 0.001\)). Origin population is
positively associated with flow size, confirming the gravity intuition
that larger countries produce more migrants. Columns (1) and (2)
additionally show that destination GDP per capita enters positively,
consistent with the brain drain hypothesis that talent flows toward
higher-income economies. Column (4) confirms these patterns hold when
the dependent variable is the number of players rather than their market
value.

\section{The Colonial Dimension}\label{sec-colonial}

The role of colonial history in shaping modern migration patterns is
well established \citep{beine2011diasporas}. Colonial ties create
linguistic, cultural, and institutional links that lower migration costs
and facilitate the acquisition of dual citizenship. In the football
context, these ties also create pathways for talent identification:
European clubs actively scout in former colonies, and players from these
regions are more likely to acquire European citizenship through ancestry
or naturalisation.

\begin{table}[!h]
\centering
\caption{\label{tab:colonial-by-coloniser}Leg drain gained by former colonisers. Total gained shows all dual-citizenship inflows; colonial value shows only flows from former colonies to their specific coloniser; share is the colonial fraction of total inflows.}
\centering
\begin{threeparttable}
\begin{tabular}[t]{lrrrrr}
\toprule
Coloniser & Colonies & Players & Total gained (€m) & From colonies (€m) & Share (\%)\\
\midrule
France & 24 & 2,116 & 3,390.5 & 2,562.3 & 75.6\\
United Kingdom & 27 & 798 & 2,898.1 & 2,035.2 & 70.2\\
Netherlands & 4 & 209 & 1,087.3 & 648.4 & 59.6\\
Portugal & 6 & 215 & 566.1 & 445.3 & 78.7\\
Spain & 13 & 60 & 692.2 & 285.7 & 41.3\\
Belgium & 2 & 122 & 535.4 & 199.1 & 37.2\\
Germany & 3 & 39 & 1,034.2 & 66.6 & 6.4\\
Italy & 2 & 12 & 397.6 & 1.0 & 0.2\\
\bottomrule
\end{tabular}
\begin{tablenotes}[para]
\item \textit{Note: } 
\item The number of colonies shown reflects those with at least one dual-citizenship player flowing to the coloniser in our data. The full colonial coding comprises 89 relationships (France: 25, United Kingdom: 28, Spain: 20, Portugal: 6, Netherlands: 4, Italy: 4, Germany: 4, Belgium: 3); colonies with no observed player flows to their specific coloniser are omitted from this table. See Appendix for complete coding.
\end{tablenotes}
\end{threeparttable}
\end{table}

Former French colonies are by far the largest source of leg drain value,
followed by British, Dutch, and Portuguese colonies. This aligns with
the composition of their national teams: France's squad routinely
includes players with roots in West Africa, the Maghreb, and the
Caribbean overseas territories. Figure A.1 in the Appendix maps these
colonial leg drain flows geographically.

\section{Discussion and Conclusion}\label{discussion-and-conclusion}

This paper introduces the concept of \emph{leg drain}---the
redistribution of football talent through dual citizenship---as a
quantifiable analogue to brain drain. Using market valuations from
Transfermarkt for 92,643 players worldwide, we document large,
systematic, and asymmetric flows of player value from developing to
developed countries.

Three results merit emphasis. First, the sheer scale of leg drain is
striking: over a third of all dual-citizenship players in our sample
represent a different national team than their country of origin,
redirecting billions of euros in player value. Second, when normalised
by population or GDP, the burden falls disproportionately on small,
low-income nations---precisely those least able to absorb the loss.
Third, colonial ties are among the strongest predictors of leg drain
intensity, suggesting that historical institutional relationships
continue to shape the global distribution of talent.

These findings speak to broader debates about globalisation and
inequality. The football labour market, while unique in many respects,
shares key features with other skilled-labour markets: talent is
globally mobile, demand is concentrated in wealthy countries, and
historical and linguistic ties shape migration corridors. The advantage
of studying football is that individual valuations are observable,
making the ``drain'' directly quantifiable in monetary terms.

Several limitations should be noted. First, Transfermarkt valuations,
while validated, are estimates rather than actual transaction prices.
Second, our dual-citizenship identification relies on the platform's
records, which may undercount players whose secondary nationality is not
officially listed. Third, our colonial-tie analysis is necessarily
coarse: we code binary relationships that in reality involve degrees of
institutional and cultural proximity. Future work could exploit more
granular measures of colonial intensity, such as duration of colonial
rule or linguistic overlap, and leverage the full market value time
series to study how leg drain evolves over time. Fourth, national-team
representation does not entail a change in club employment or residence;
leg drain captures the redistribution of sporting talent rather than
productive labour. The bilateral patterns we document are nonetheless
shaped by the same structural forces --- colonial ties, linguistic
links, diaspora networks --- that drive labour migration.

Despite these caveats, the leg drain framework offers a transparent,
reproducible method for studying the global redistribution of human
capital. As football continues to globalise and dual citizenship becomes
more common, the phenomenon is likely to intensify---raising important
questions about equity, representation, and the governance of
international sport.

\subsection*{Declaration of generative AI and AI-assisted technologies
in the manuscript preparation
process}\label{declaration-of-generative-ai-and-ai-assisted-technologies-in-the-manuscript-preparation-process}
\addcontentsline{toc}{subsection}{Declaration of generative AI and
AI-assisted technologies in the manuscript preparation process}

During the preparation of this work the authors used Claude (Anthropic)
in order to assist with manuscript drafting and typesetting in
Quarto/LaTeX. After using this tool/service, the authors reviewed and
edited the content as needed and take full responsibility for the
content of the published article.

\section*{References}\label{references}
\addcontentsline{toc}{section}{References}

\renewcommand{\bibsection}{}
\bibliography{references}

\appendix

\section*{Appendix}\label{appendix}
\addcontentsline{toc}{section}{Appendix}

\subsection*{Bilateral Flows}\label{bilateral-flows}
\addcontentsline{toc}{subsection}{Bilateral Flows}

\begin{figure}[H]

{\centering \includegraphics{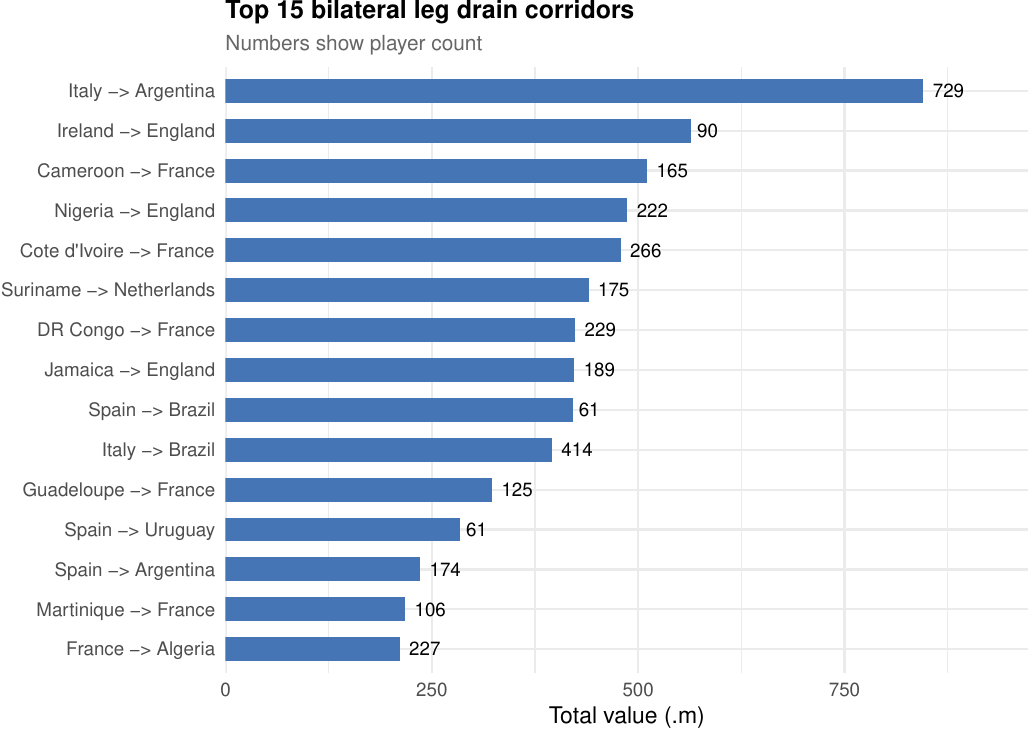}

}

\caption{Top 15 bilateral leg drain corridors by total value (€
millions). Arrow direction: from origin (losing) to destination
(gaining) country.}

\end{figure}

\subsection*{Colonial Ties: Map and
Coding}\label{colonial-ties-map-and-coding}
\addcontentsline{toc}{subsection}{Colonial Ties: Map and Coding}

\begin{figure}[H]

{\centering \includegraphics{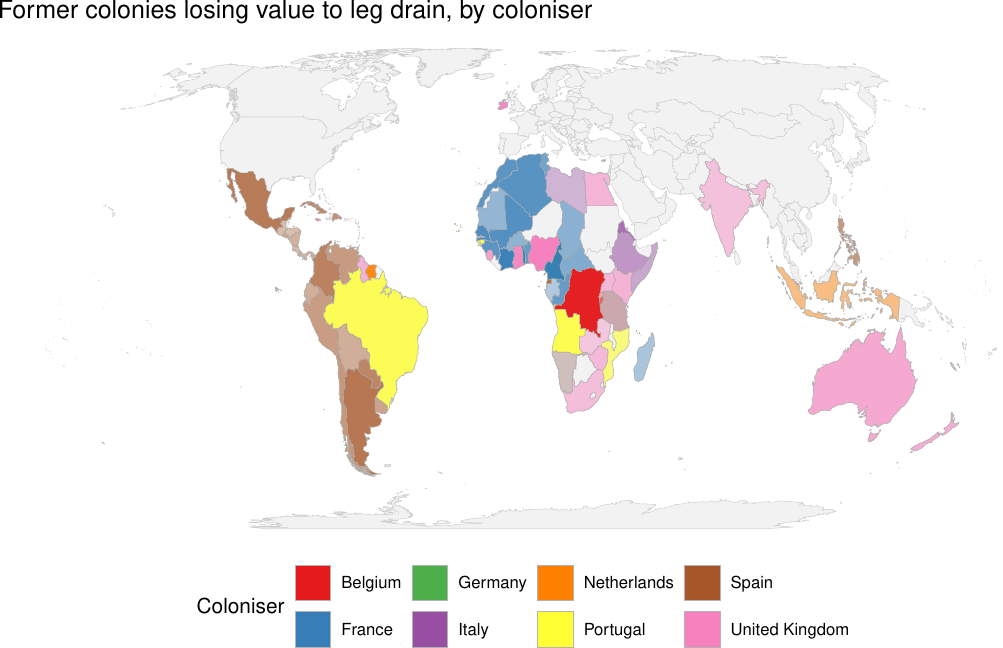}

}

\caption{Value lost to leg drain by former colonies, coloured by
coloniser. Darker shades indicate higher value lost.}

\end{figure}

Our colonial-tie coding covers 89 bilateral relationships across eight
European colonisers, drawing on the CEPII GeoDist database
\citep{mayer2011geodist}, the ICOW Colonial History Data Set
\citep{hensel2018icow}, and \citet{head2010erosion}. The United Kingdom
accounts for 28 former colonies, followed by France (25), Spain (20),
Portugal (6), Netherlands (4), Italy (4), Germany (4), and Belgium (3).
Four territories---Cameroon, Tanzania, Namibia, and Togo---are
dual-coded under two colonisers: each was initially a German colony
(1884--1919) before passing to France or Britain under League of Nations
mandates after the Treaty of Versailles. In the bilateral gravity model,
these dual-coded territories create separate observations for each
coloniser pair. Germany's colonial ties are the weakest in our coding,
lasting only roughly 30 years; modern institutional, linguistic, and
migration links to Germany are minimal, and results are robust to
excluding German colonial ties entirely. Current French overseas
departments (Guadeloupe, Martinique, French Guiana, Mayotte, Réunion,
New Caledonia) and British overseas territories (Bermuda, Montserrat)
are also coded as colonies, since players from these territories
represent a direct colonial pathway to the metropole's national team.

\end{document}